\documentclass[aps,prd, nofootinbib,showpacs]{revtex4-1}
\usepackage{graphicx,epsf,amssymb,url,hyperref}
\usepackage{amsmath,amsfonts,amssymb}
\usepackage{bm}
\usepackage{mathrsfs}
\usepackage[]{latexsym}
\usepackage[UKenglish]{babel}
\usepackage{epsfig}
\usepackage[utf8]{inputenc}
\usepackage{soul}
\usepackage{amsthm}
\usepackage{color}
\newcommand{\be}{\begin{eqnarray}}
\newcommand{\ee}{\end{eqnarray}}

\newcommand{\lp}{\ell_{\rm p}}
\newcommand{\mpl}{m_{\rm p}}

\newcommand{\rh}{r_{\rm H}}

\newcommand{\Lambdac}{\Lambda_{cutoff}}

\def\comment#1{}

\definecolor{darkred}{rgb}{.8,0,0}

\definecolor{darkblue}{rgb}{0,0,.7}

\definecolor{darkgreen}{rgb}{0,.7,0}

\begin{document}

%
%
%%%%%%%%%%%%%%%%%%%%%%%%%%%%%%%%%%%%%%%%%%%%%%%%%%%%%%%%%%%%%%
\title{The GUP and quantum Raychaudhuri equation} 
%%%%%%%%%%%%%%%%%%%%%%%%%%%%%%%%%%%%%%%%%%%%%%%%%%%%%%%%%%%%%%
%

%
%
%
%
\author{Elias~C.~Vagenas$^1$}\email[email:~]{elias.vagenas@ku.edu.kw}
\author{Lina~Alasfar$^2$}\email[email:~]{ lina.alasfar@clermont.in2p3.fr}
\author{Salwa~M.~Alsaleh$^3$}\email[email:~]{salwams@ksu.edu.sa}
\author{Ahmed~Farag~ Ali$^{4,5}$} \email[email:~]{ahmed.faragali@nias.knaw.nl}
\affiliation{$^1$Theoretical Physics Group, Department of Physics, Kuwait University, P.O. Box 5969, Safat 13060, Kuwait}
\affiliation{$^2$Laboratoire de Physique  de Clermont-Ferrand, Universit\'{e} Clermont Auvergne, 4, Avenue Blaise Pascal 	63178 Aubi\`{e}re Cedex, France }
\affiliation{$^3$Department of Physics and Astronomy, King Saud University, Riyadh 11451, Saudi Arabia}
\affiliation{$^4$Netherlands Institute for Advanced Study, Korte Spinhuissteeg 3, 1012 CG Amsterdam, Netherlands}
\affiliation{$^5$Department of Physics, Faculty of Science, Benha University, Benha, 13518, Egypt}
%
%
%
%
%\date{\today}
%
%%%%%%%%%%%%
\begin{abstract}
%%%%%%%%%%%%
%
\par\noindent
In this paper, we compare the quantum corrections to the Schwarzschild black hole temperature due to 
quadratic and linear-quadratic generalized uncertainty principle, with the corrections from the quantum 
Raychaudhuri equation. The reason for this comparison is to connect the deformation parameters  
$\beta_0$ and  $ \alpha_0$ with  $\eta$ which is the parameter that characterizes the quantum 
Raychaudhuri equation. The derived relation between the parameters appears to depend 
on the relative scale of the system (black hole), which could be read as a beta function
equation for the quadratic deformation parameter $\beta_0$. This study shows a correspondence between the two phenomenological approaches and indicates that quantum Raychaudhuri equation implies the existence of a 
crystal-like structure of spacetime. 
\end{abstract}

\maketitle
%
%%%%%%%%%%%%%%%
%%%%%%%%%%%%%%%
% 
%
%
\par\noindent
One of the most challenging questions in theoretical physics over the last century is to obtain 
a fully consistent  theory of quantum gravity. Although many attempts to reach such theory have been made, 
namely loop quantum gravity, string theory and other approaches, it is believed that such programmes 
are not yet complete and have their own problems. Hence, it appears that it is important to 
investigate quantum gravity  from a phenomenological perspective~\cite{Hossenfelder:2009nu}. 
In the last few decades, many phenomenological approaches have been developed to investigate 
quantum gravity at the Plank scale~\cite{Hossenfelder:2006mi}. 
\par\noindent
One of these phenomenological approaches is the generalisation of Heisenberg uncertainty 
principle~(HUP), which is inconsistent with general relativity, as  HUP implies that it is impossible to 
localise a system without adding a lot of energy to it. However, the energy of this measurement will affect 
the background space-time according to general relativity, making the smooth structure of the classical 
space-time impossible. The generalised uncertainty principle~(GUP) resolves  this inconsistency by changing 
the HUP such that it has a minimum length  or, equivalently, a maximum momentum. The generalisation of 
HUP has been studied in many contexts~\cite{GUPearly}. The most prominent space deformations of 
Heisenberg algebra, that imply GUP, consist of either quadratic momentum term~\cite{VenezGrossMende} 
\be
\left[\hat{X},\hat{P}\right] = i\,\hbar\left(1 + \beta_0\, \frac{\hat{P}^{2}}{\mpl^{2}}\right),
\label{[1]}
\ee
with $\beta_0$ to be the dimensionless deforming parameter of GUP, or, both linear and quadratic terms~\cite{Ali},
\be
[\hat X,\hat P] = i \hbar \left( 1- 2\alpha \hat P +  4\alpha^2 \hat P^2 \right)
\label{2}
\ee
where $\alpha= \alpha_0/\mpl=\alpha_0\ell_p/\hbar$ with $\alpha_0$ to be the dimensionless GUP parameter 
and  $\mpl$ to be the Planck mass (we have set $c=1$).
\par\noindent
The quadratic deformations were derived from doubly special relativity~(DSR), loop quantum gravity, 
and string theory ~\cite{VenezGrossMende,MM,kempf,FS,Adler2,SC}, while the linear-quadratic 
deformations were developed later as a generalisation~\cite{Ali}. The GUP has  important implications 
in cosmology and black hole physics~\cite{alsaleh}. Moreover, it has indicated 
a cyclic non-singular universe ~\cite{GUPcosmology},  black hole remnants  and also that 
the deformed black hole thermodynamics derived from GUP appears to coincide with the one  
obtained from studying black hole entropy in loop quantum gravity and string theory \cite{GUPBH}. 
Furthermore, it has been shown that the GUP produces a logarithmic correction 
to entropy similar to that derived on  statistical grounds~\cite{Medved:2004yu}. 
\par \noindent
 The above-mentioned results  derived from the GUP, also  appear in other phenomenological 
approaches (cf.~\cite{Das:2009hs}). The most recent of which is the quantum Raychaudhuri
 equation~(QRE)~\cite{Das:2013oda}
\begin{equation}
\dot \theta = \frac{1}{3} \theta ^2 - \sigma^2 +\Omega^2 - R_{\mu \nu} \xi^\mu \xi^\nu 
 -\nonumber \frac{\epsilon_1 \hbar^2}{m^2}h^{\mu \nu} R_{; \mu \nu}- \frac{\hbar^2}{m^2}h^{\mu \nu }
\left( \frac{\square \mathcal R}{\mathcal R}\right) _{;\mu \nu}~.
\label{QRE}
\end{equation}
The QRE is derived from considering the pilot wave theory of quantum mechanics to assign to 
each particle moving in the congruence a semi-classical, WKB-type wavefunction
~$ \psi (x) = \mathcal R e^{-i \mathcal S x}$. The  phenomenological implications of~QRE 
to  cosmology~\cite{Ali:2014qla} and black holes~\cite{Ali:2015tva} were very similar 
to the ones obtained from GUP. On one hand, the deformed black hole temperature can be obtained 
either from the quadratic GUP and  it takes the form~\cite{Scardiglia}
\be
T_{QGUP} = \frac{\pi}{\beta_0} \left(  M-\sqrt{M^2-\frac{\beta_0}{\pi^2}m_p^2}\,\right) ,
\label{TQGUP}
\ee
or, from the linear-quadratic GUP and it becomes~\cite{alsaleh2}
%
%
%
%
%\begin{widetext}
\be
T_{LQGUP} = \frac{\left(4\pi \displaystyle\hbar\frac{M}{m_p^2}  + \displaystyle\frac{\hbar\alpha_0}
{\mpl}\right)-\sqrt{\left(4\pi \displaystyle\hbar\frac{M}{m_p^2}+
\displaystyle \frac{\hbar\alpha_0}{\mpl}\right)^{2}
-4\left(\displaystyle\frac{4\hbar\alpha_0^2}{\mpl^{2}}\right) \hbar}}
	{2\left(\displaystyle\frac{4\hbar\alpha_0^2}{\mpl^{2}}\right)}~.
\label{TLQGUP}
\ee
%\end{widetext}
%
%
%
%
On the other hand, the deformed black hole temperature can be obtained from studying the quantum-corrected 
Schwarzschild metric from QRE and it takes the form~\cite{Ali:2015tva}
\be
T_{QRE} = \hbar \frac{\sqrt{M^2-4\eta m_p^2 }}{2 \pi  \left(\sqrt{M^2-4\eta m_p^2 }+M\right)^2}
\ee 
with  $\eta$ to be the QRE parameter.
\par
\begin{figure}
	\centering
	\includegraphics[width= 0.7 \linewidth]{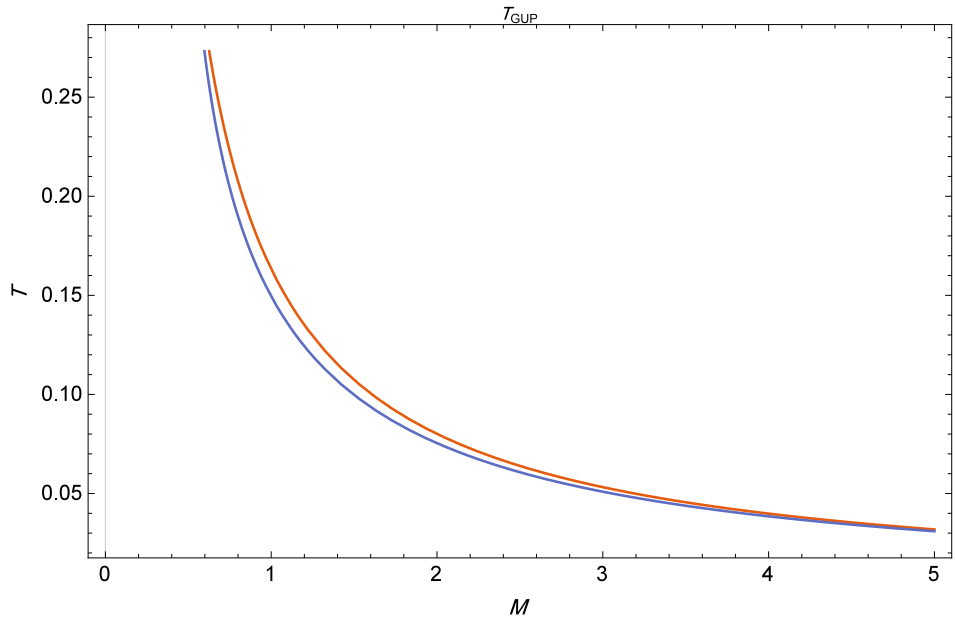}
	\caption{ Plot of quantum-corrected black hole temperature by quadratic GUP (orange) and linear-quadratic GUP (blue) vs the black hole mass. Here we considered~$ m_p  \sim 1$ and~$\alpha_0 \sim 1$. }
	\label{TGUP}
\end{figure}

\begin{figure}
	\centering
	\includegraphics[width= 0.7 \linewidth]{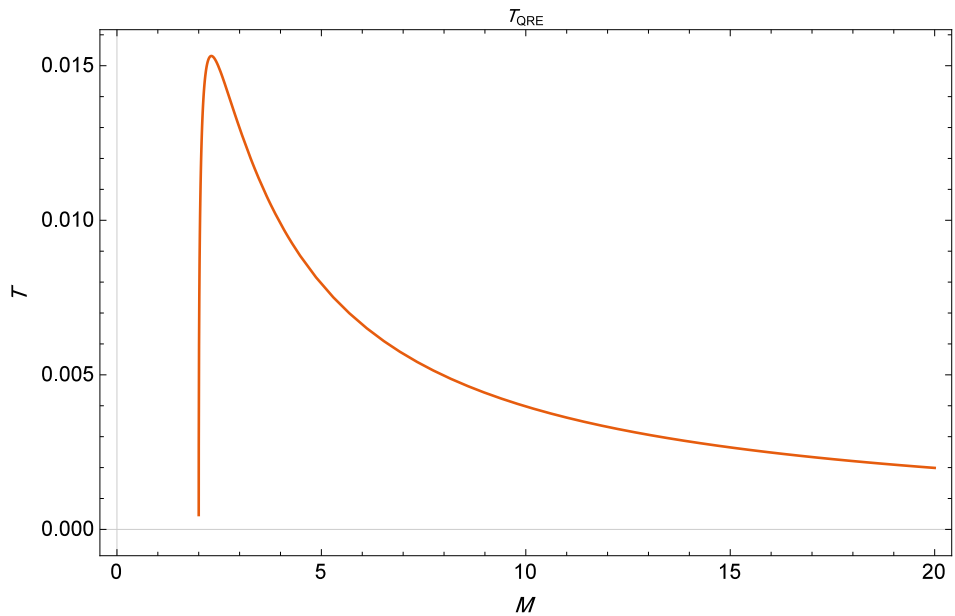}
	\caption{A plot of quantum-corrected black hole temperature by QRE vs the black hole mass. Here we considered~$ m_p \eta \sim 1$. }
	\label{TQRE}
\end{figure}

\noindent
It is evident that both deformations predict the existence of a black hole remnant of 
minimal mass~$M_{min}$ and prevent the black hole from a catastrophic evaporation, 
(see figures~\ref{TGUP} and~\ref{TQRE}).
This could suggest that both approaches describe  same physics and one may transform 
one approach into the other. In order to investigate this would-be correspondence, we will expand the expressions of  the 
deformed black hole temperatures and compare their subleading terms to establish,  if this is possible, a relation between 
GUP and QRE.
\par\noindent
Next, we argue that the deformed QRE metric would imply a quantum correction 
to Newton's law, similar to the one obtained from the tree diagrams of graviton exchange in the weak 
field limit~\cite{Duff:1974ud,Dono}. It has already been shown that the GUP corrections have a thermal 
nature ~\cite{Scardiglia}, and that both quadratic and linear-quadratic GUP deformations correspond to the tree-level 
correction to the Schwarzschild metric~\cite{Scardiglia,alsaleh2,Ali:2014dfa}. 
To explicitly show the above correspondence between QRE and GUP, we need to 
relate the two deformation parameters. Thus, we start with the QRE-corrected Schwarzschild metric~\cite{Ali:2015tva}
\be
ds^2 = \left( 1-\frac{2 M}{r}+\frac{4\eta\, \mpl^2}{r^2}\right)dt^2 - 
\left( 1-\frac{2 M}{r}+\frac{4\eta \,\mpl^2}{r^2}\right)^{-1} dr^2-r^2 d\Omega^2
\label{QREmetric}
\ee
which implies the quantum-corrected gravitational potential in the weak field limit 
(since~$V(r) = \frac{1}{2}(g_{tt}-1 )$)
\be
V(r) \simeq -\frac{M}{r}\left( 1-2\eta \, \mpl\,\frac{\mpl }{ Mr}\right) ~.
\label{potential}
\ee
It is easily seen that  in the RHS of  (\ref{potential})  the second term which is the QRE correction term to the potential in the weak field limit, is scale-dependent since this term contains the relative scale of the system, i.e., $ \mpl/M$.
Thus, since the GUP corrections are not scale-dependent, we expect the relation between the QRE parameter $ \eta$ and the dimensionless GUP parameter $\alpha_0$ to be also scale-dependent.
We now rewrite metric~\eqref{QREmetric} as
\be
ds^2 = \left( 1-\frac{2 M}{r}+\varepsilon(r)\right)dt^2 -
\left( 1-\frac{2 M}{r}+\varepsilon(r)\right)^{-1} dr^2-r^2 d\Omega^2
\label{QREm2}
\ee
where~$ \varepsilon(r) \ll M/r$. Therefore, the horizon equation takes the form
\be
1- \frac{2M}{r} + \varepsilon(r) =0,
\ee
and its solution reads
\be
r_{h}= M+\sqrt{M^2-4 \eta \, \mpl^2}~.
\ee
Furthermore, the black hole temperature is given by the relation~\cite{Townsend:1997ku}
\vspace{-2ex}
\begin{align}
T=&  \frac{\hbar}{4 \pi}\,\lim_{r\to\rh}\left[ g_{tt,r}(r) \right] \nonumber \\
=& \hbar \lim_{r\to\rh} \frac{2 M+r^2 \epsilon '(r)}{4 \pi  r^2} \\
=& \hbar  \lim_{r\to\rh}\frac{M r-4 \eta \, \mpl^2}{2 \pi  r^3}\nonumber~.
\end{align}
By taking the limit~$ r\to r_h$, we obtain the expression for the QRE deformed temperature
\be
T_{QRE}= \hbar\frac{M}{2 \pi  \left(\sqrt{M^2-4 \eta  \mpl^2}+ M\right)^2}+\frac{\hbar\epsilon '(r_h)}{4 \pi }.
\hspace{2ex}
\ee
%
%
%
%
%\par\noindent
Then, considering black holes with $M >> \mpl$, we expand  the above QRE  
deformed temperature and obtain
\be
T_{QRE} \simeq \frac{\hbar}{8 \pi M}  \left[ 1- \eta^2 \,L^4- 2 \eta^3 L^6+ \dots\right]
\label{expandTQRE}
\ee
where~$ L = \mpl/M$ is  the relative scale of the system. 
\par \noindent
At this point, we expand the quadratic GUP-corrected black hole temperature given by \eqref{TQGUP}, 
in terms of the relative scale of the system, up to second order,\footnote{One can ignore higher 
order terms since the GUP  corrections are most trusted at first and second order  \cite{Ali:2011fa}.} 
\be
T_{QGUP} = \frac{\hbar}{8 \pi M}\left[ 1+ \frac{1}{4 \pi ^2} \beta_{0} L^2+ \dots\right] ,
\label{expansionTQGUP}
\ee
%
%
%
%
%\par\noindent
as well as the linear-quadratic GUP-corrected black hole temperature given by \eqref{TLQGUP}, 
also up to second order,
 \be 
 T_{LQGUP}= \frac{\hbar}{8 \pi M}\left[ 1- \frac{\alpha_0}{2 \pi}L+ 5 \left( \frac{\alpha_0}
{2 \pi}\right) ^2 L^2 + \dots \right]~.\hspace{2ex}
\label{expansionTLQGUP}
 \ee
 %
 %
%\par \noindent
Now, we can compare the subleading terms of \eqref{expandTQRE} 
with the subleading terms of the expansion of the quadratic GUP given by \eqref{expansionTQGUP} 
and obtain
 \be
 \eta^2 = - \frac{1}{4 \pi^2} \beta_{0} L^{-2}~.
 \label{beta}
\ee
\par\noindent
In addition, we can compare the subleading terms of \eqref{expandTQRE} 
with the subleading terms of   \eqref{expansionTLQGUP} and obtain
 \be
 \eta^2 = \frac{\alpha_0}{2 \pi}L^{-3}-5\left( \frac{\alpha_0}{2 \pi}\right) ^2 L^{-2}~.
 \label{beta1}
 \ee
At this point a number of comments are in order.
\par\noindent
First, as expected from our comments below (\ref{potential}), the QRE parameter, i.e., $\eta$, depends on the relative scale of the system, i.e., $L=\mpl/M$. Additionally, both relations, namely  \eqref{beta} and \eqref{beta1},  reveal a deep connection between GUP and QRE. 
Furthermore, these relations can be obtained even if higher order terms in the GUP-deformed temperatures were included. 
\par\noindent
Second, it was recently shown that the dimensionless GUP parameter  $\beta_0$ cannot be negative 
since it implies a minimum length of $\Delta x \ge \hbar\sqrt{\beta_0}$ which will be imaginary 
if $\beta_0$ is negative \cite{Ali:2015zua}.
However, it was shown in Refs. \cite{Jizba:2009qf, Scardigli:2014qka} that one can consider a negative 
GUP parameter  $\beta_0$ if the uncertainties are computed on a crystal-like universe and the 
lattice spacing of the universe is of the order of Planck length, i.e. $\lp$. 
Therefore, the negativity of $\beta_0$ in (\ref{beta}) can be interpreted as a signal of a lattice structure of space-time. Consequently, this  gives the possibility of having a geometric interpretation of the QRE, namely that the Bohmian trajectories of geodesics may be produced by a crystal-like structure of space-time, i.e.,  a discrete geometry may give Bohmian trajectories. 
%
%
%\textcolor{blue}{This is consistent with the stochastic interpretation of quantum mechanics~\cite{sto,Zakir:1999gg} }
%
%
%
%
%
%
%
%
\par\noindent
Third,  (\ref{beta})  is a relation between the two parameters, namely QRE and GUP parameters,  which can also be viewed as a series expansion in powers of the relative scale $L$ of the system.  From this perspective,  in
the context of statistical mechanics as well as quantum field theory, this relation can be read as  the beta function  equation for the effective  coupling  constant $\beta_0$ with the energy scale to be  the black hole mass, i.e., $ \Lambda = M$, 
and the cut-off scale to be  the Planck mass, i.e., $\Lambdac\sim \mpl$.
%
%
%
%.
 %
 %
 %
It should be stressed that  in Ref. \cite{Faizal:2017dlb} the quadratic deformation parameter, i.e., $\beta_0$, was found to play the r\^{o}le of the coupling constant in an effective field theory. 
%
%
%
%
%
%
%
%
%%%%%%%%%%%%%%%%%%%%%%%%%%%
\section*{Acknowledgements}
%%%%%%%%%%%%%%%%%%%%%%%%%%%
%
%
%
\vspace{-2ex}
\par\noindent
We would like to thank Saurya Das for useful correspondences.
This research project was supported by a grant from the  
``\textit{Research Center of the Female Scientific and Medical Colleges}", 
Deanship of Scientific Research, King Saud University.
%
%
%
%
%%%%%%%%%%%%%%%%%

%%%%%%%%%%%%
%
%
%

\begin{thebibliography}{99}
%%%%%%%%%%%%%%%%
%
%
%
%
%
\bibitem{Hossenfelder:2009nu} 
 S.~Hossenfelder and L.~Smolin,
  %``Phenomenological Quantum Gravity,''
  Phys.\ Canada {\bf 66}, 99 (2010)
  [arXiv:0911.2761 [physics.pop-ph]].
%
%
%
%
\bibitem{Hossenfelder:2006mi} 
S.~Hossenfelder,
  %``Phenomenological quantum gravity,''
  AIP Conf.\ Proc.\  {\bf 903}, 463 (2007)
  doi:10.1063/1.2735224
  [hep-th/0611017].
%
%
%
%
%
%
\bibitem{GUPearly}
H.~S.~Snyder,
  %``Quantized space-time,''
  Phys.\ Rev.\  {\bf 71}, 38 (1947),
  doi:10.1103/PhysRev.71.38;
 C.~N.~Yang,
  %``On quantized space-time,''
  Phys.\ Rev.\  {\bf 72}, 874 (1947),
  doi:10.1103/PhysRev.72.874;
 C.~A.~Mead,
  %``Possible Connection Between Gravitation and Fundamental Length,''
  Phys.\ Rev.\  {\bf 135}, B849 (1964),
  doi:10.1103/PhysRev.135.B849;
 F.~Karolyhazy,
  %``Gravitation and quantum mechanics of macroscopic objects,''
  Nuovo Cim.\ A {\bf 42}, 390 (1966),
  doi:10.1007/BF02717926.
%
%
%
%
%
%
%
\bibitem{VenezGrossMende}
D.~Amati, M.~Ciafaloni and G.~Veneziano,
  %``Superstring Collisions at Planckian Energies,''
  Phys.\ Lett.\ B {\bf 197}, 81 (1987), 
  doi:10.1016/0370-2693(87)90346-7;
D.~J.~Gross and P.~F.~Mende,
  %``The High-Energy Behavior of String Scattering Amplitudes,''
  Phys.\ Lett.\ B {\bf 197}, 129 (1987), 
  doi:10.1016/0370-2693(87)90355-8;
D.~Amati, M.~Ciafaloni and G.~Veneziano,
  %``Can Space-Time Be Probed Below the String Size?,''
  Phys.\ Lett.\ B {\bf 216}, 41 (1989), 
  doi:10.1016/0370-2693(89)91366-X;
K.~Konishi, G.~Paffuti and P.~Provero,
  %``Minimum Physical Length and the Generalized Uncertainty Principle in String Theory,''
  Phys.\ Lett.\ B {\bf 234}, 276 (1990), 
  doi:10.1016/0370-2693(90)91927-4;
G.~Veneziano,
  ``Quantum string gravity near the Planck scale,''
  CERN-TH-5889-90, C90-03-27; 
S.~Capozziello, G.~Lambiase and G.~Scarpetta,
  %``Generalized uncertainty principle from quantum geometry,''
  Int.\ J.\ Theor.\ Phys.\  {\bf 39}, 15 (2000)
  doi:10.1023/A:1003634814685
  [gr-qc/9910017].
%
%
%
%
%
%
%
\bibitem{Ali} 
  A.~F.~Ali, S.~Das and E.~C.~Vagenas,
  %``Discreteness of Space from the Generalized Uncertainty Principle,''
  Phys.\ Lett.\ B {\bf 678}, 497 (2009),
  doi:10.1016/j.physletb.2009.06.061
  [arXiv:0906.5396 [hep-th]].
%
%
%
%
\bibitem{MM}
 M.~Maggiore,
  %``A Generalized uncertainty principle in quantum gravity,''
  Phys.\ Lett.\ B {\bf 304}, 65 (1993),
  doi:10.1016/0370-2693(93)91401-8
  [hep-th/9301067].
%
%
%
%
\bibitem{kempf}
 A.~Kempf, G.~Mangano and R.~B.~Mann,
  %``Hilbert space representation of the minimal length uncertainty relation,''
  Phys.\ Rev.\ D {\bf 52}, 1108 (1995),
  doi:10.1103/PhysRevD.52.1108
  [hep-th/9412167];
M.~Bojowald and A.~Kempf,
  %``Generalized uncertainty principles and localization of a particle in discrete space,''
  Phys.\ Rev.\ D {\bf 86}, 085017 (2012),
  doi:10.1103/PhysRevD.86.085017
  [arXiv:1112.0994 [hep-th]].
%
%
%
%
\bibitem{FS}
 F.~Scardigli,
  %``Generalized uncertainty principle in quantum gravity from micro - black hole Gedanken experiment,''
  Phys.\ Lett.\ B {\bf 452}, 39 (1999), 
  doi:10.1016/S0370-2693(99)00167-7
  [hep-th/9904025].
%
%
%
%
\bibitem{Adler2}
  R.~J.~Adler and D.~I.~Santiago,
  %``On gravity and the uncertainty principle,''
  Mod.\ Phys.\ Lett.\ A {\bf 14}, 1371 (1999), 
  doi:10.1142/S0217732399001462
  [gr-qc/9904026].
%
%
%
\bibitem{SC}
 F.~Scardigli and R.~Casadio,
  %``Generalized uncertainty principle, extra dimensions and holography,''
  Class.\ Quant.\ Grav.\  {\bf 20}, 3915 (2003),
  doi:10.1088/0264-9381/20/18/305
  [hep-th/0307174].
%
%
%
%
%
\bibitem{alsaleh}
B.~Majumder,
  %``Effects of GUP in Quantum Cosmological Perfect Fluid Models,''
  Phys.\ Lett.\ B {\bf 699}, 315 (2011),
  doi:10.1016/j.physletb.2011.04.030
  [arXiv:1104.3488 [gr-qc]];
F.~Darabi and K.~Atazadeh,
  ``GUP, Einstein static universe and cosmological constant problem,''
  arXiv:1704.03040 [gr-qc]; 
 K.~Zeynali, F.~Darabi and H.~Motavalli,
  %``Multi-Dimensional Cosmology and DSR-GUP,''
  Mod.\ Phys.\ Lett.\ A {\bf 28}, 1350047 (2013),
  doi:10.1142/S0217732313500478
  [arXiv:1212.1571 [gr-qc]];
S.~Alsaleh, A.~Al-Modlej and A.~F.~Ali,
  ``Virtual Black Holes from Generalized Uncertainty Principle and Proton Decay,''
  arXiv:1703.10038 [physics.gen-ph].
%
%
%
%
%
\bibitem{GUPcosmology} 
M.~Salah, F.~Hammad, M.~Faizal and A.~F.~Ali,
  %``Non-singular and Cyclic Universe from the Modified GUP,''
  JCAP {\bf 1702}, no. 02, 035 (2017),
  doi:10.1088/1475-7516/2017/02/035
  [arXiv:1608.00560 [gr-qc]].
%
%
%
%
%
\bibitem{GUPBH}
 S.~Gangopadhyay, A.~Dutta and A.~Saha,
  %``Generalized uncertainty principle and black hole thermodynamics,''
  Gen.\ Rel.\ Grav.\  {\bf 46}, 1661 (2014),
  doi:10.1007/s10714-013-1661-3
  [arXiv:1307.7045 [gr-qc]];
  R.~J.~Adler, P.~Chen and D.~I.~Santiago,
  %``The Generalized uncertainty principle and black hole remnants,''
  Gen.\ Rel.\ Grav.\  {\bf 33}, 2101 (2001),
  doi:10.1023/A:1015281430411
  [gr-qc/0106080];
M.~Faizal and M.~M.~Khalil,
  %``GUP-Corrected Thermodynamics for all Black Objects and the Existence of Remnants,''
  Int.\ J.\ Mod.\ Phys.\ A {\bf 30}, no. 22, 1550144 (2015),
  doi:10.1142/S0217751X15501444
  [arXiv:1411.4042 [gr-qc]].
%%
%
%
%
%
%
%
%
\bibitem{Medved:2004yu} 
A.~J.~M.~Medved and E.~C.~Vagenas,
  %``When conceptual worlds collide: The GUP and the BH entropy,''
  Phys.\ Rev.\ D {\bf 70}, 124021 (2004),
  doi:10.1103/PhysRevD.70.124021
  [hep-th/0411022].
%
%
%
%%
%
%
\bibitem{Das:2009hs} 
S.~Das and E.~C.~Vagenas,
  %``Phenomenological Implications of the Generalized Uncertainty Principle,''
  Can.\ J.\ Phys.\  {\bf 87}, 233 (2009),
  doi:10.1139/P08-105
  [arXiv:0901.1768 [hep-th]];
 A.~F.~Ali, S.~Das and E.~C.~Vagenas,
 ``The Generalized Uncertainty Principle and Quantum Gravity Phenomenology,''
doi:10.1142/9789814374552$\_$0492
arXiv:1001.2642 [hep-th].
%
%
%
%
%
%
%
%
%
\bibitem{Das:2013oda} 
 S.~Das,
  %``Quantum Raychaudhuri equation,''
  Phys.\ Rev.\ D {\bf 89}, no. 8, 084068 (2014),
  doi:10.1103/PhysRevD.89.084068
  [arXiv:1311.6539 [gr-qc]].
%
%
%
%

%
%
\bibitem{Ali:2014qla} 
A.~F.~Ali and S.~Das,
 A.~F.~Ali and S.~Das,
  %``Cosmology from quantum potential,''
  Phys.\ Lett.\ B {\bf 741}, 276 (2015),
  doi:10.1016/j.physletb.2014.12.057
  [arXiv:1404.3093 [gr-qc]].
%
%
%
%
%
%
\bibitem{Ali:2015tva} 
A.~F.~Ali and M.~M.~Khalil,
  %``Black Hole with Quantum Potential,''
  Nucl.\ Phys.\ B {\bf 909}, 173 (2016),
  doi:10.1016/j.nuclphysb.2016.05.005
  [arXiv:1509.02495 [gr-qc]].
%
%
%
%
%
\bibitem{Scardiglia} 
F.~Scardigli, G.~Lambiase and E.~Vagenas,
  %``GUP parameter from quantum corrections to the Newtonian potential,''
  Phys.\ Lett.\ B {\bf 767}, 242 (2017), 
  doi:10.1016/j.physletb.2017.01.054
  [arXiv:1611.01469 [hep-th]].
%
%
%
%
%
\bibitem{alsaleh2}
 E.~C.~Vagenas, S.~M.~Alsaleh and A.~Farag,
  %``GUP parameter and black hole temperature,''
  EPL {\bf 120}, no. 4, 40001 (2017)
  doi:10.1209/0295-5075/120/40001
  [arXiv:1801.03670 [hep-th]].
%
%
%
%
\bibitem{Duff:1974ud}
 M.~J.~Duff,
  %``Quantum corrections to the schwarzschild solution,''
  Phys.\ Rev.\ D {\bf 9}, 1837 (1974),
  doi:10.1103/PhysRevD.9.1837.
%
%
%
%
%
%
\bibitem{Dono}
J.~F.~Donoghue,
  %``Leading quantum correction to the Newtonian potential,''
  Phys.\ Rev.\ Lett.\  {\bf 72}, 2996 (1994),
  doi:10.1103/PhysRevLett.72.2996
  [gr-qc/9310024];
J.~F.~Donoghue,
  %``General relativity as an effective field theory: The leading quantum corrections,''
  Phys.\ Rev.\ D {\bf 50}, 3874 (1994)
  doi:10.1103/PhysRevD.50.3874
  [gr-qc/9405057].
%
%
%
%
%
\bibitem{Ali:2014dfa} 
 A.~F.~Ali and M.~Moussa,
  %``Towards Thermodynamics with Generalized Uncertainty Principle,''
  Adv.\ High Energy Phys.\  {\bf 2014}, 629148 (2014),
  doi:10.1155/2014/629148.
%
%
%
%
%\cite{Townsend:1997ku}
\bibitem{Townsend:1997ku} 
  P.~K.~Townsend,
 ``Black holes: Lecture notes,''
  gr-qc/9707012.
  %%CITATION = GR-QC/9707012;%%
%
%
%
%
%
%
%\cite{Ali:2011fa}
\bibitem{Ali:2011fa}
  A.~F.~Ali, S.~Das and E.~C.~Vagenas,
  %``A proposal for testing Quantum Gravity in the lab,''
  Phys.\ Rev.\ D {\bf 84}, 044013 (2011),
  doi:10.1103/PhysRevD.84.044013
  [arXiv:1107.3164 [hep-th]].
%
%

%
%
%
%\cite{Ali:2015zua}
\bibitem{Ali:2015zua} 
  A.~Farag Ali, M.~M.~Khalil and E.~C.~Vagenas,
  %``Minimal Length in quantum gravity and gravitational measurements,''
  Europhys.\ Lett.\  {\bf 112}, no. 2, 20005 (2015), 
  doi:10.1209/0295-5075/112/20005
  [arXiv:1510.06365 [gr-qc]].
%
%
%
%
%
%
%\cite{Jizba:2009qf}
\bibitem{Jizba:2009qf} 
  P.~Jizba, H.~Kleinert and F.~Scardigli,
  %``Uncertainty Relation on World Crystal and its Applications to Micro Black Holes,''
  Phys.\ Rev.\ D {\bf 81}, 084030 (2010),
  doi:10.1103/PhysRevD.81.084030
  [arXiv:0912.2253 [hep-th]].
%
%
%
%
%
%\cite{Scardigli:2014qka}
\bibitem{Scardigli:2014qka} 
F.~Scardigli and R.~Casadio,
  %``Gravitational tests of the Generalized Uncertainty Principle,''
  Eur.\ Phys.\ J.\ C {\bf 75}, no. 9, 425 (2015),
  doi:10.1140/epjc/s10052-015-3635-y
  [arXiv:1407.0113 [hep-th]].
%
%
%
%\bibitem{sto}
%C.~Frederick,
  %``Stochastic Space-Time and Quantum Theory,''
 %Phys.\ Rev.\ D {\bf 13}, 3183 (1976).
  %doi:10.1103/PhysRevD.13.3183
%
%
%
%
%\bibitem{Zakir:1999gg} 
%Z.~Zakir,
  %``The Theory of stochastic space-time. 2. Quantum theory of relativity,''
 % Theor.\ Phys.\ Astrophys.\ Cosmol.\  {\bf 4}, 10 (2009), 
 % doi:10.9751/TPAC.3200-014
 % [hep-th/9901013].
%
%
%
%
\bibitem{Faizal:2017dlb} 
  M.~Faizal, A.~F.~Ali and A.~Nassar,
  %``Generalized uncertainty principle as a consequence of the effective field theory,''
  Phys.\ Lett.\ B {\bf 765}, 238 (2017),
  doi:10.1016/j.physletb.2016.11.054
[arXiv:1701.00341 [hep-th]].
%
%
%
%

%%%%%%%%%%%%%%%%%%%%%%%%%%%%%%%%%%%%%%%%%%%%%%
%
%
%
%%%%%%%%%%%%
\end{thebibliography}
\end{document}